\documentclass[12pt]{iopart}

\usepackage{amssymb}
\usepackage{amstext}
\usepackage{graphicx}
\usepackage{color}
\usepackage{ulem}

\newcommand {\be} {\begin{equation}}
\newcommand {\ee} {\end{equation}}
\newcommand {\bq} {\begin{eqnarray}}
\newcommand {\eq} {\end{eqnarray}}

\def\etal{{\it et al.}}

\begin{document}

\title{Quantum Peierls stress of straight and kinked dislocations and effect of non-glide stresses}

\author{B. Barvinschi$^{1,2}$, L. Proville$^3$, D. Rodney$^1$}

\address{$^1$Science et Ing\'enierie des Mat\'eriaux et Proc\'ed\'es, Grenoble Institute of Technology, UJF, CNRS, Saint Martin d'H\`{e}res 38402, France\\
$^2$Faculty of Physics, West University of Timisoara, Timisoara 300223, Romania\\
$^3$CEA, DEN, Service de Recherches de M\'etallurgie Physique, Gif-sur-Yvette 91191, France}
\ead{david.rodney@grenoble-inp.fr}
\begin{abstract}
It was recently shown \cite{proville-natmat2012} that to predict reliable Peierls stresses from atomistic simulations, one has to correct the Peierls barrier by the zero-point energy difference between the initial and activated states of the dislocation. The corresponding quantum Peierls stresses are studied here in $\alpha$-Fe modeled with two embedded atom method potentials. First, we show that the quantum correction arises from modes localized near the dislocation core, such that partial Hessian matrices built on small cylinders centered on the dislocation core can be used to compute the zero-point energy difference. Second, we compute quantum Peierls stresses for straight and kinked dislocations and show that the former is smaller than the latter with both $\alpha$-Fe models. Finally, we compare quantum Peierls stresses obtained in simple shear and in traction along two orientations considered experimentally by Kuramoto \etal~\cite{kuramoto-pm1979}, evidencing a strong effect of non-glide stresses on the quantum Peierls stress.
\end{abstract}

\maketitle

%%%%%%%%%%%%%%%%%%%%%%%%%%%%%%%%%%%%%%%%%%%%%%%%%%%%%%%%%%%%%%%%%%%%%%%%%%%%%%%%%%%%%%%
\section{Introduction}
\label{sec:intro}
%%%%%%%%%%%%%%%%%%%%%%%%%%%%%%%%%%%%%%%%%%%%%%%%%%%%%%%%%%%%%%%%%%%%%%%%%%%%%%%%%%%%%%%

The Peierls stress of a dislocation \cite{peierls-ppsl1940,rodney-physmetal} measures its lattice friction, that is the intrinsic resistance of the crystal to the motion of the dislocation; motion which proceeds through the nucleation and propagation of kink-pairs, a process known as the Peierls mechanism. It has long been recognized that measurements of the Peierls stress from atomistic models systematically overestimate experimental data. This discrepancy was first reported by Basinski \textit{et al.} \cite{basinski-cjp1971} in their early calculations in sodium modeled with a pair potential. But the same effect has since been repeatedly reported, in particular in body-centered cubic (BCC) metals using more advanced energetic models, including embedded atom method (EAM) potentials \cite{wen-am2000,chaussidon-am2006,gordon-msmse2010}, bond-order potentials \cite{groger-am2008a,mrovec-prl2011,chen-msmse2013} and even ab initio density functional theory (DFT) \cite{woordward-prl2002,Ventelon2007,ventelon-am2013,weinberger-prb2013}.

Simulating the process of kink-pair formation requires rather large simulation cells, which is currently prohibitive for instance for DFT calculations. But to compute the Peierls stress, we can take advantage of the fact that in this limit, the critical kink pair has zero length and the dislocation glides as a straight line. The translational symmetry along the dislocation then allows to reduce the cell size in this direction to a single Burgers vector, which drastically reduces the number of atoms. Moreover, it was shown using EAM potentials \cite{rodney-prb2009,proville-prb2013}, that classical Peierls stresses for 
straight and kinked dislocations are strictly equivalent. The discrepancy between experimental data and atomistic simulations is therefore not an artifact of the computational approach but has instead a physical origin.

It was shown recently \cite{proville-natmat2012} that the difference between experimental and simulated Peierls stresses is mainly due to a quantum effect 
arising from the zero-point motion of the atoms near the dislocation core. This effect has been considered in several simplified dislocation models \cite{Suzuki1970,Gilman1968,Takeuchi1982} but had so far been systematically discarded in atomistic simulations. This effect is however expected even in heavy metals as Fe because the Debye temperature, which marks the onset of quantum effects in the vibrational modes of the system, is high (470 K in Fe) and the experiments to measure the Peierls stress were performed down to very low temperatures, 4 K and below \cite{kuramoto-pm1979,brunner-zfm1992}.

The theory employed in Ref. \cite{proville-natmat2012} recovers at low temperature the usual zero-point correction of activated processes, first derived by Wigner \cite{Wigner1938}: in the limit of zero temperature, the classical enthalpy barrier at a given applied stress, $\sigma$,  is decreased by the zero-point energy difference between initial and activated states:
\begin{equation}
\Delta H_{quantum} = \Delta H_{classic} - \Delta E_{ZP}
\end{equation}
with
\begin{equation}
\Delta E_{ZP} = \sum_i h \nu^{init}_i / 2 - \sum_k h \nu^{\star}_k / 2, 
\label{eq:correction}
\end{equation}
where $\{\nu_i^{init} \}$ and $\{\nu_k^{\star} \}$ are the real non-zero eigenfrequencies computed from the positive eigenvalues of the Hessian matrix diagonalized in the initial and activated states at the applied stress $\sigma$. The quantum Peierls stress, $\sigma_P^Q$, may then be defined by $\Delta H_{quantum}\left(\sigma_P^Q \right) = 0$, while the classical Peierls stress, $\sigma_P^C$, corresponds to $\Delta H_{classic}\left(\sigma_P^C \right) = 0$.

The calculations presented in Ref. \cite{proville-natmat2012} were performed in BCC Fe in three-dimensional (3D) cells containing a kinked $1/2 \langle 111 \rangle$ screw dislocation. Since developing realistic interatomic potentials in BCC metals is notoriously difficult, two EAM potentials, developed by Gordon \textit{et al.} \cite{Gordon2010} and Marinica \cite{proville-natmat2012}, were compared. In both cases, the quantum correction $\Delta E_{ZP}$ was found large, on the order of 0.1 eV, leading to a quantum Peierls stress near 450 MPa with Gordon potential, only slightly larger than the accepted experimental value, 400 MPa, while the quantum Peierls stress was higher with Marinica potential, $\sim$ 600 MPa.

The above Peierls stresses were obtained loading the simulation cell in simple shear where the only non-zero component is the resolved shear stress (RSS), i.e. the shear stress along the $1/2 \langle 111 \rangle$ Burgers vector resolved on the $(\bar{1}01)$ glide plane. Experimental data on the other hand are based on uniaxial tractions that necessarily involve non-glide stresses, i.e. stress components that produce no Peach-Koehler force on the dislocation and yet, are known to strongly influence dislocation glide in BCC crystals \cite{duesbery-prsl1984,ito-pma2001,groger-am2008a,chen-msmse2013}. A consequence is that the Peierls stress, expressed as a critical resolved shear stress (CRSS), depends on the orientation of the crystal with respect to the traction axis. For instance, in the experimental work of Kuramoto \textit{et al.} \cite{kuramoto-pm1979}, two orientations, noted $A$ and $B$, were considered and lead to two different Peierls stresses, respectively 375 MPa and 450 MPa. In the present paper, we will compare quantum Peierls stresses obtained in simple shear with that obtained for the two orientations considered by Kuramoto \textit{et al.}, in order to account for and assess the influence of non-glide stresses. Interestingly, this will require to compute the effect of non-glide stresses on the full kink-pair formation process and not just on the Peierls stress of straight dislocations as done in the past \cite{duesbery-prsl1984,ito-pma2001,groger-am2008a,chen-msmse2013}.

Also, computing the quantum correction requires in principle to diagonalize the full Hessian matrix of the system, which represents a computational challenge since 3D cells with acceptable finite size effects contain at least 100,000 atoms. However, it was shown in Ref. \cite{proville-natmat2012} that the main contribution to the quantum correction comes from modes localized near the dislocation core, such that 90 $\%$ of the correction is recovered considering a partial Hessian matrix computed in a cylinder of radius 7 $\AA$ around the dislocation core. We will study here more systematically the dependence of the quantum correction on the cylinder radius in order to determine an optimum radius to perform the diagonalizations.

Finally, even when the calculations are limited to a small cylinder, the number of atoms remains too large ($\sim 15,000$) to be treated from first principles. 
This is regretful since ab inito calculations could provide a more quantitative estimate of the quantum Peierls stress than semi-empirical potentials. 
One way to drastically reduce the number of atoms is to compute the quantum correction on a straight dislocation in a simulation cell reduced to a single Burgers vector along the dislocation line. As mentioned above, Peierls stresses computed for straight and kinked dislocations are identical in the classical limit, but in the quantum regime, the Peierls stress  
is reached while the dislocation is still kinked in the activated state. It is thus not clear whether the quantum Peierls stress computed for a straight dislocation, 
where the full dislocation is brought up the Peierls barrier, is consistent with the quantum Peierls stress of a kinked dislocation where 
only the kink regions deviate 
from the bottom of the Peierls valley.

The aim of the present work is thus three fold: (1) study systematically the dependence of the quantum correction on the cylinder radius, (2) compare quantum 
Peierls stresses of straight and kinked dislocations, (3) compute quantum Peierls stresses in traction conditions to account for the non-glide stresses 
met experimentally.

%%%%%%%%%%%%%%%%%%%%%%%%%%%%%%%%%%%%%%%%%%%%%%%%%%%%%%%%%%%%%%%%%%%%%%%%%%%%%%%%%%%%%%%
\section{Methodology}
\label{sec:methodo}
%%%%%%%%%%%%%%%%%%%%%%%%%%%%%%%%%%%%%%%%%%%%%%%%%%%%%%%%%%%%%%%%%%%%%%%%%%%%%%%%%%%%%%%

%%%%%%%%%%%%%%%%%%%%%%%%%%%%%%%%%%%%%%%%%%%%%%%%%%%%%%%%%%%%%%%%%%%%%%%%%%%%%%%%%%%%%%%
\subsection{Simulation cell and interatomic potentials}
\label{sec:sim_cell}
%%%%%%%%%%%%%%%%%%%%%%%%%%%%%%%%%%%%%%%%%%%%%%%%%%%%%%%%%%%%%%%%%%%%%%%%%%%%%%%%%%%%%%%

\begin{figure}
\begin{center}
\includegraphics[width=10cm]{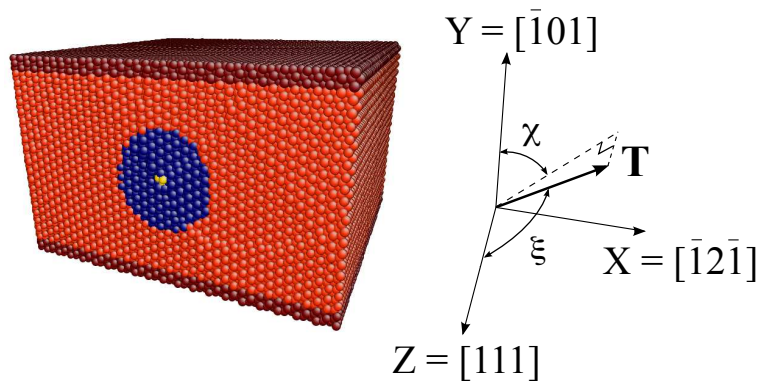}
\caption{\label{fig:simcell}Simulation cell and corresponding crystallographic axes. Yellow atoms belong to the dislocation core, blue atoms show an example of cylinder with radius $R = 20 \AA$ used to compute partial Hessian matrices, dark red atoms near the upper and lower $Y$ surfaces are used to apply external stresses in traction-controlled boundary conditions.}
\end{center}
\end{figure}

The simulation cell and associated crystallography are shown in Fig. \ref{fig:simcell}. The cell directions are $X = [\bar{1} 2 \bar{1}]$, $Y = [\bar{1} 0 1]$ and $Z = [ 1 1 1]$. We introduce a screw dislocation with a $b=1/2[111]$ Burgers vector along the central $Z$ axis of the cell by means of its elastic displacement field. Periodic boundary conditions are applied in the $X$ and $Z$ directions, while free surfaces are used in the $Y$ direction. The nominal cell dimensions are $L_X= 138~\AA$, $L_Y= 98~\AA$, $L_Z = 132~\AA$ for kinked dislocations and $L_Z = 2.4~\AA = b$ for straight dislocations. A shift $\delta = b/2$ is added in the $Z$ direction to the periodic boundary conditions across the $X$ surfaces in order to account for the plastic strain introduced by the screw dislocation. More details about the simulation cell are given in Refs. \cite{rodney-am2004,bacon-dislinsolids}.

We employed two EAM interatomic potentials, developed by Gordon \textit{et al.} \cite{Gordon2010} and Marinica \cite{proville-natmat2012}. Both potentials were specifically adjusted to model dislocations in BCC Fe and predict a compact non-degenerate easy core structure and a $(\bar{1} 0 1)$ glide plane in simple shear. One usual shortcoming of interatomic potentials with a non-degenerate core is that they  predict a metastable split core \cite{Takeuchi1979,gordon-msmse2010,ventelon-am2013}, resulting in bent dislocation paths between easy cores passing through the split core \cite{ventelon-am2013} and Peierls potentials with an intermediate minimum \cite{gordon-msmse2010}, in contrast with DFT calculations where the paths avoid the split core and the Peierls barriers have a single maximum \cite{Ventelon2007,ventelon-am2013,weinberger-prb2013}. Both Gordon and Marinica potentials were developed to avoid this artifact. With Gordon potential, the split core is still metastable but the local minimum in the Peierls potential is very shallow, while with Marinica potential, the split core is unstable and the Peierls potential has a single maximum. The quantum correction however depends on the second derivatives of the potentials that are not included in the fitting procedure. In Ref. \cite{proville-natmat2012}, Gordon potential was shown to predict more realistically phonon spectra than Marinica potential, but the same may not be true for the quantum correction, which arises from variations of localized vibration modes in the course of dislocation activation. 

%In the following, the predictions of both potentials will be compared, in both 2D and 3D, in order to draw at least qualitative conclusions concerning the quantum correction. 

%%%%%%%%%%%%%%%%%%%%%%%%%%%%%%%%%%%%%%%%%%%%%%%%%%%%%%%%%%%%%%%%%%%%%%%%%%%%%%%%%%%%%%%
\subsection{Boundary conditions for general applied stress tensor}
\label{sec:BC}
%%%%%%%%%%%%%%%%%%%%%%%%%%%%%%%%%%%%%%%%%%%%%%%%%%%%%%%%%%%%%%%%%%%%%%%%%%%%%%%%%%%%%%%

If a uniaxial stress $\sigma$ is applied along a traction axis $\overrightarrow{T}$ (unit vector), the corresponding stress tensor is

\begin{equation}
\overline{\overline{\sigma}} = \sigma \overrightarrow{T} \otimes \overrightarrow{T},
\end{equation}
where $\otimes$ stands for the tensor (outer) product. As recalled in Fig. \ref{fig:simcell}, traction axes are usually referred to by the angle $\xi$ between $\overrightarrow{T}$ and the $[111]$ axis (the $Z$ axis here) and the angle $\chi$ between $[\bar{1} 0 1 ]$ and the projection of $\overrightarrow{T}$ in the $(111)$ plane. The angle $\chi$ measures the orientation of the maximum resolved shear stress plane (MRSSP) with respect to the $(\bar{1} 0 1)$ glide plane. Expressing the components of $\overrightarrow{T}=\left[T_X T_Y T_Z \right]$ along $X$, $Y$ and $Z$, as a function of $\xi$ and $\chi$, we have:

\begin{eqnarray}
\overline{\overline{\sigma}} &=& \sigma \left[ \begin{array}{ccc} T_X^2&T_X T_Y&T_X T_Z\\
&T_Y^2&T_Y T_Z\\
&&T_Z^2 \end{array}
\right] \nonumber \\  &=&
\sigma \left[ \begin{array}{ccc}
\sin^2\xi \sin^2 \chi&
\frac{1}{2} \sin^2 \xi \sin 2 \chi&
\frac{1}{2} \sin 2 \xi \sin \chi\\
&\sin^2 \xi \cos^2 \chi&
\frac{1}{2} \sin 2 \xi \cos \chi\\
&&\cos^2 \xi
\end{array}
\right].
\label{eq:stress_tensor}
\end{eqnarray}

Such a general stress tensor can be applied to the present simulation cell by a combination of Andersen-type extended system equations in the periodic directions \cite{allen-1987} and traction-controlled boundary conditions in the $Y$ direction where free surfaces are used \cite{bacon-dislinsolids}. Andersen-type barostats are used to control $\sigma_{XX}$, $\sigma_{ZZ}$ and $\sigma_{XZ}$ by applying equations of motion to the cell dimensions $L_X$, $L_Z$ and the shift $\delta$. For $L_X$ for instance, we integrate:
\begin{equation}
M \ddot{L}_X = \left( \sigma_{XX} - \sigma^{app}_{XX} \right) L_Y L_Z,
\end{equation}
where $M$ is a numerical mass, $\sigma_{XX}$ the current stress and $\sigma^{app}_{XX}$ the applied stress. Since only quasi-static simulations are performed, we employed a quenched molecular dynamics algorithm to remove the thermal energy. Similar equations were applied to relax $\sigma_{ZZ}$ through $L_Z$ and $\sigma_{XZ}$ through $\delta$.

The same approach cannot be used for the stress components involving the $Y$ direction because of the free surfaces. Instead we used traction-controlled boundary conditions by adding external forces to the atoms within the potential cutoff radius from the upper and lower $Y$ surfaces (dark red atoms in Fig. \ref{fig:simcell}). Forces in direction $I$ ($I=X, Y, Z$) are added to control the stress component $\sigma_{YI}$. The forces are of opposite signs in the two surfaces and are scaled in order to apply the desired stress. For more details, see Ref. \cite{bacon-dislinsolids}.

Usually, with simple shear, the cell dimensions are held fixed, such that the stress components $\sigma_{XX}$, $\sigma_{ZZ}$ and $\sigma_{XZ}$ are not zero (on the order of a few 100 MPa). In the following we relax these stresses to be consistent between loading conditions but we checked that in the case of simple shear, they have no effect on dislocation glide.

Duesbery \cite{duesbery-prsl1984} showed that the tensor in Eq. \ref{eq:stress_tensor} can be usefully decomposed into a hydrostatic pressure, a glide tensor that produces the resolved shear stress (RSS) on the dislocation and two non-glide tensors, called the isotropic and edge tensors, that produce no RSS. Among the latter, only the glide and edge tensors were shown to affect dislocation mobility \cite{duesbery-prsl1984,ito-pma2001,groger-am2008a}. The glide tensor produces the RSS, which is simply $\sigma_{YZ}$ expressed as $RSS = \frac{1}{2} \sigma \sin 2 \xi \cos \chi$. The edge tensor corresponds to a pure shear perpendicular to the dislocation Burgers vector, which we will call non-glide stress (NGS), of magnitude $NGS = \frac{1}{2} \sigma \sin ^2 \xi \cos 2 \chi$. In simulations, the cells are usually loaded in simple shear, where only the RSS is different from zero, while in experiments performed in traction or compression, the non-glide stress is usually non-zero. In order to evaluate how the latter affects the quantum Peierls stress, we will consider the two orientations used in the work of Kuramoto \etal~\cite{kuramoto-pm1979}, with  $\chi = -1^\text{o}$ and $\xi = 46^\text{o}$ for orientation $A$ and $\chi = 29^\text{o}$ and $\xi = 37^\text{o}$ for orientation $B$.

\subsection{Minimum enthalpy paths}
\label{sec:MEP}

The initial and final configurations, with the dislocation located in two successive Peierls valleys, were loaded quasistatically using the above combination of Andersen-type and force-controlled boundary conditions. Since the initial and final configurations differ only by a translation of the dislocation in the periodic $X$ direction, the cell shape, which varies with applied stress, remains the same in both configurations during loading.

Minimum enthalpy paths between these configurations were obtained using the nudged elastic band (NEB) method as implemented in Refs. \cite{rodney-prb2007,rodney-prb2009,proville-natmat2012}. The paths were discretized with 60 images. Straight dislocations were generated by interpolating atomic positions linearly between initial and final configurations. Kinked dislocations were created by taking the atomic positions from the initial configuration except in a slab along the $Z$ direction where the positions were taken from the final configuration. Increasing linearly the width of the slab generates a path with a dislocation containing an expanding kink pair. After an initial relaxation, the path was iteratively re-relaxed near its maximum in order to determine the activated state with a force threshold of 10$^{-4}$ eV/$\AA^2$, required for accurate eigenvalue spectra. The NEB calculations were performed at fixed cell shape, i.e. we removed the Andersen barostats and kept only the external forces, whose work was included in the potential energy of the system. This simplification is justified by the fact that the cell shape is the same in initial and final configurations and is expected to vary only very slightly along the paths, with negligible influence on the activated state.

Once the activated state is identified, the activation enthalpy is simply the difference in enthalpy between activated and initial states and the quantum correction in the limit of zero temperature, i.e. the difference in zero-point energy between initial and activated states (see Eq. \ref{eq:correction}) is obtained by diagonalizing the Hessian matrix of the system in both states, using either all the atoms in the system or only the atom within a cylinder around the dislocation core. Exact diagonalizations were performed using the Math Kernel Library \cite{intel}.

\section{Dependencies of the quantum correction}
\label{sec:analQC}

\subsection{Influence of the cylinder radius}
\label{sec:cylinder}

\begin{figure}
\begin{center}
\includegraphics[width=14cm]{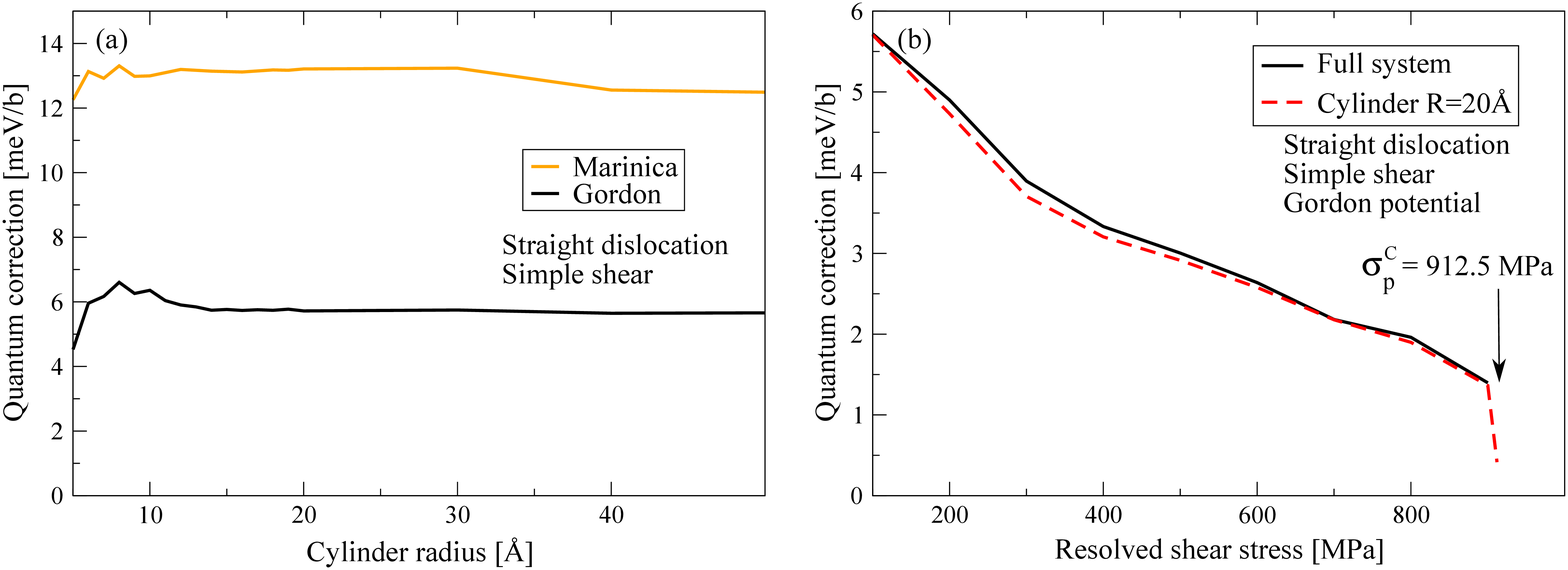}
\caption{\label{fig:cylinder2D} Dependence of the quantum correction on the cylinder radius for a straight dislocation loaded in simple shear: (a) as a function of cylinder radius for an applied stress of 100 MPa for both Gordon and Marinica potentials, (b) as a function of applied stress for Gordon potential computed either in a cylinder of radius 20 $\AA$ or in the full system.}
\end{center}
\end{figure}

\begin{figure}
\begin{center}
\includegraphics[width=7cm]{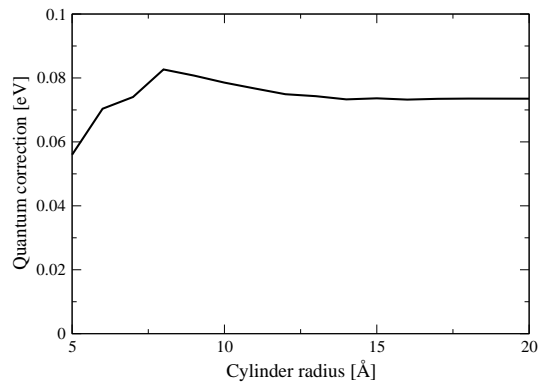}
\caption{\label{fig:cylinder3D}Dependence of the quantum correction on the cylinder radius for a kinked dislocation loaded in simple shear at 400 MPa with Gordon potential.}
\end{center}
\end{figure}

The dependence of the quantum correction on the cylinder radius was studied for both straight and kinked dislocations. The results with straight dislocations are summarized in Fig. \ref{fig:cylinder2D}. Fig. \ref{fig:cylinder2D}(a) shows the quantum correction as a function of the cylinder radius for Gordon and Marinica potentials. Down to a radius of about 15 $\AA$, the correction is essentially independent of the cylinder radius with Gordon potential and varies by less than 5 $\%$ with Marinica potential. Fig. \ref{fig:cylinder2D}(b) compares the quantum correction computed for a straight dislocation 
in a cylinder of radius $R = 20$ $\AA$ and in the full system as a function of applied stress, showing that the difference is less than 10 $\%$ over the entire stress range. Calculations with a kinked dislocation
are shown in Fig. \ref{fig:cylinder3D}. As in previous case, the quantum correction is essentially independent of the cylinder radius down to a radius on the order of 12 $\AA$ and compares very well with the correction obtained for the full system in Ref. \cite{proville-natmat2012}, 0.07 eV. The correction then varies by less than 15 $\%$ down to a radius of 5 $\AA$.

These calculations confirm that for both straight and kinked dislocations, the quantum correction arises from modes localized near the dislocation core. 
In the following, for kinked dislocations, we will perform diagonalizations in cylinders of radius 20 $\AA$, which is the largest radius that does not require high performance computing, 
in order to account for the largest possible volume around the dislocation core but we should keep in mind that, to perform faster calculations, radii down to about 10 $\AA$ can be used as shown in \ref{fig:cylinder3D}. For straight dislocations, we could also use a reduced cylinder but because the calculations are fast in this geometry, we performed the diagonalizations on the full systems.

\subsection{Mode analysis}
\label{sec:mode}

\begin{figure}
\begin{center}
\includegraphics[width=14cm]{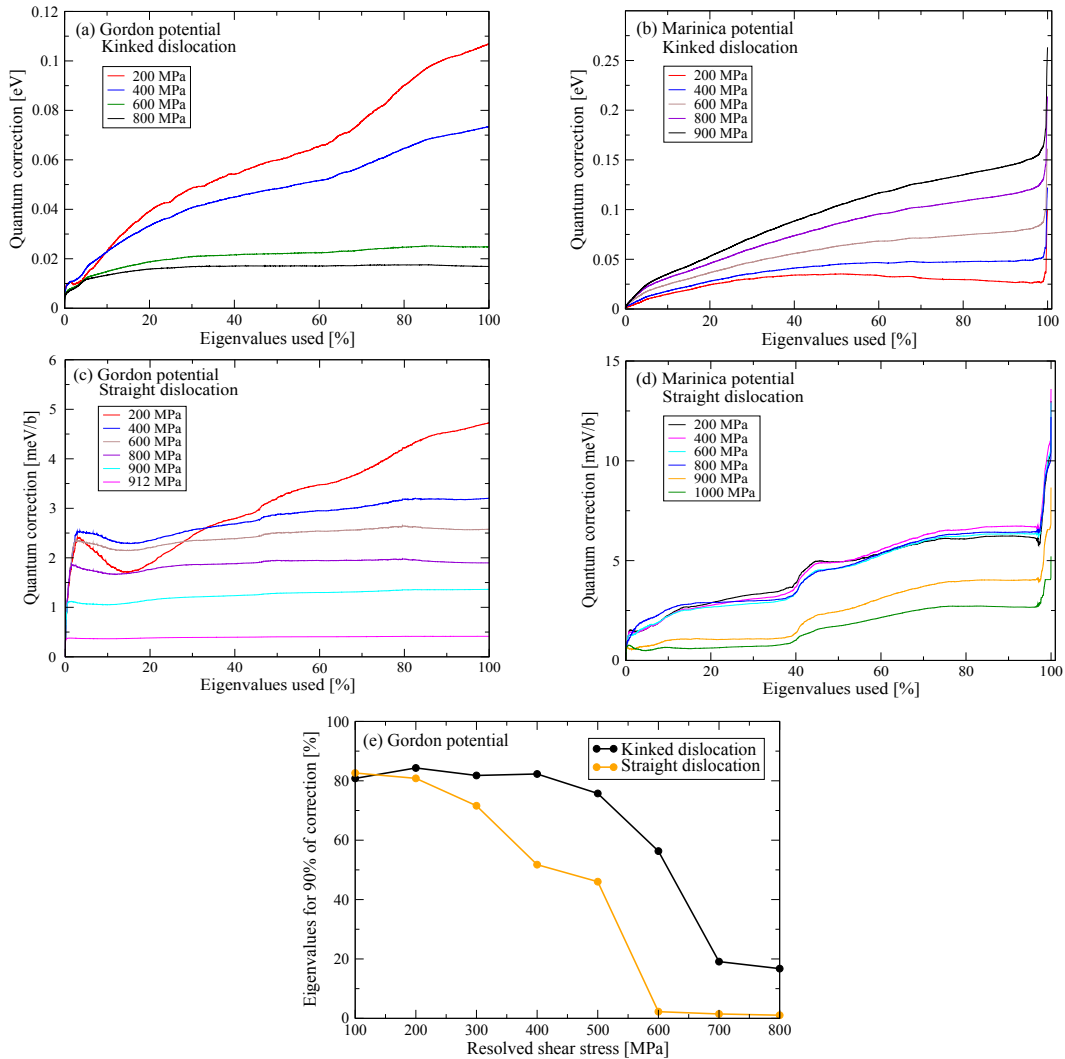}
\caption{\label{fig:mode}Quantum correction as a function of the fraction of modes used to computed Eq. \ref{eq:correction} for straight and kinked dislocations with Gordon and Marinica potentials. In (e), the fraction of modes required to account for 90 $\%$ of the quantum correction is reported as a function of applied stress with Gordon potential.}
\end{center}
\end{figure}

To gain more insights on the origin of the quantum correction, we analyzed its dependence on the number of modes used to evaluate Eq. \ref{eq:correction}. To this end, after diagonalization of the Hessian matrix, we sorted the eigenmodes in ascending order of eigenvalue, or equivalently of eigenfrequency, and computed the quantum correction from Eq. \ref{eq:correction} limiting the summations to a given number of modes. Thus, if $M$ modes are used, the correction involves the $M$ (resp. $M-1$) first non-zero eigenvalues of the initial (resp. activated) configuration. The result is shown in Fig. \ref{fig:mode} for both straight and kinked
dislocations modeled with Gordon and Marinica potentials.

Considering Gordon potential first, we see from Fig. \ref{fig:mode}(a) that for a kinked dislocation, the quantum correction increases gradually with the number of modes. As shown Fig. \ref{fig:mode}(e), the fraction of modes required to account for 90 $\%$ of the correction decreases with the applied stress but remains on the order of 20 $\%$ even at high stresses. This behavior, already reported in Ref. \cite{proville-natmat2012}, implies that the correction, while arising from modes localized near the dislocation core, cannot be ascribed to a single or a few modes but rather to the accumulation of small differences over a large number of modes.

The situation is different for a straight dislocation, where, as seen in Fig. \ref{fig:mode}(c), the correction increases very rapidly with the first few modes and then much more gradually across the rest of the spectrum. As a result, the fraction of modes required to account for 90 $\%$ of the correction decreases much more rapidly with stress for a straight dislocation than for a kinked dislocation, 
as shown in Fig. \ref{fig:mode}(e). In the limit of high stresses near the classical Peierls stress ($\sim$ 912.5 MPa for Gordon potential), the entire quantum correction arises from a very small number of low-frequency modes. In this limit, the mode mostly responsible for the correction is the soft mode in the initial configuration whose eigenvalue vanishes at the classical Peierls stress. The reason is that the system undergoes at the Peierls stress a fold instability (also called a saddle-node bifurcation \cite{guckenheimer-book}) where both a positive eigenvalue in the initial configuration and the negative eigenvalue in the activated state vanish. As a consequence, the quantum correction necessarily vanishes at the classical Peierls stress, as confirms Fig. \ref{fig:cylinder2D}(b) where the correction drops down to zero at high stresses. However we see that the drop occurs over a very narrow range of stress, between about 900 and 912 MPa. The influence of the fold instability is therefore limited to the immediate vicinity of the transition. We also checked that the eigenvector of the vanishing mode reflects the atomic displacements involved at the beginning of the transition. Interestingly, the same phenomenology was observed in amorphous solids under quasistatic condition of plastic deformation \cite{maloney-pre2006,bible}.

Surprizingly, Marinica potential has a very different behavior. As seen in Fig. \ref{fig:mode}(b) and (d), a significant fraction of the quantum correction is due to a few very high frequency modes both for straight
and kinked dislocations that could correspond to optical modes in the dislocation core. Moreover, the quantum correction for a kinked dislocation with Marinica potential increases with applied stress while it decreases with Gordon potential. Finally the quantum correction for a straight dislocation is almost independent of the applied stress until close to the classical Peierls stress while it decreases progressively with Gordon potential. The reason for these strong differences between Gordon and Marinica potentials, which are both based on the formalism proposed by Mendelev \etal~\cite{mendelev-pm2003}, is unknown for the moment but emphasizes the necessity to compare between several potentials and to favor DFT-based calculations.

\section{Quantum Peierls stresses}

\subsection{Straight dislocation quantum Peierls stress}

\begin{figure}
\begin{center}
\includegraphics[width=14cm]{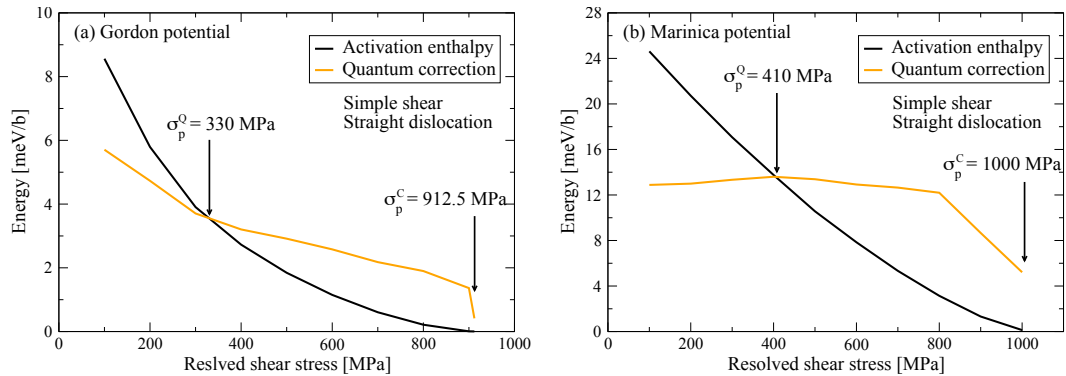}
\caption{\label{fig:2DPeierls}Activation enthalpy and quantum correction of a straight dislocation loaded in simple shear for (a) Gordon and (b) Marinica potentials. Diagonalizations were performed on the full systems.}
\end{center}
\end{figure}

Calculations in simple shear on straight dislocations are shown in Fig. \ref{fig:2DPeierls} for Gordon and Marinica potentials. 
Both the classical activation enthalpy and the quantum correction are shown as a function of applied stress. In these graphs, the quantum Peierls stress corresponds to when the quantum correction 
crosses the enthalpy curve, while the classical Peierls stress is when the enthalpy itself vanishes. Despite the differences shown in previous Section, both potentials behave in a similar manner at least qualitatively, with a large quantum correction on the order of half the kink-pair activation enthalpy at zero applied stress and a quantum Peierls stress two to three times smaller than the classical Peierls stress.

Quantitatively however, the potentials differ on several aspects. First, activation enthalpies are larger with Marinica potential than with Gordon potential. The reason is that Marinica potential predicts a higher kink-pair activation enthalpy at zero applied stress
\cite{proville-natmat2012,proville-prb2013} due to its adjustment on the ab initio value (27 meV$/$b) \cite{Ventelon2007}, which is much larger than predicted by Gordon potential (12 meV$/$b). However, the higher enthalpies with Marinica potential come along with larger quantum corrections (partly due to the large contribution of high frequency modes described in previous Section), such that quantum Peierls stresses are comparable with both potentials, 330 MPa with Gordon potential and 410 MPa with Marinica potential. Another difference is that, as discussed in previous Section, the quantum correction with Marinica potential is mostly independent of the applied stress up to close to the classical Peierls stress, while with Gordon potential, the correction decreases gradually until it drops to zero near the classical Peierls stress. This behavior is markedly different from the activation enthalpy, which decreases gradually with increasing applied stress for both potentials.

\begin{figure}
\begin{center}
\includegraphics[width=14cm]{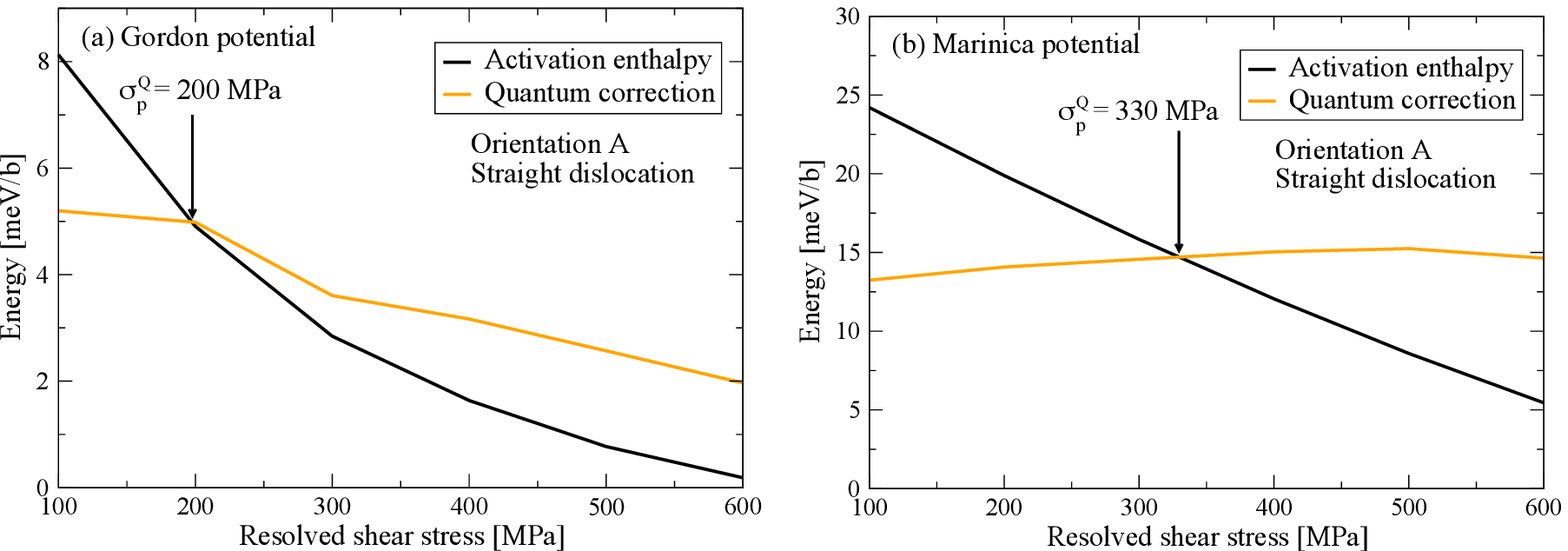}
\caption{\label{fig:2DKuramoto}Activation enthalpy and quantum correction of a straight dislocation loaded along orientation $A$ ($\chi = -1^o$, $\xi = 46^o$) for (a) Gordon and (b) Marinica potentials.}
\end{center}
\end{figure}

Fig. \ref{fig:2DKuramoto} considers a cell loaded in traction along the orientation $A$ of Kuramoto \etal~\cite{kuramoto-pm1979}. In this orientation, $\chi = -1^\text{o}$, meaning that the MRSSP remains close to the $(\bar{1}01)$ glide plane. The difference with simple shear is that there are non-glide stresses with a relatively large ratio $NGS/RSS = 0.52$. As seen by comparison between Fig. \ref{fig:2DPeierls} and Fig. \ref{fig:2DKuramoto}, these non-glide stresses strongly decrease both the classical and quantum Peierls stresses. For Gordon potential for instance, the classical Peierls stress decreases from 912 MPa  to 650 MPa while the quantum Peierls stress decreases from 330 MPa to 200 MPa, a remarkably low value.

With orientation $B$, the activation enthalpy and quantum correction are similar to Figs. \ref{fig:2DPeierls} and \ref{fig:2DKuramoto} and are not shown here. In this orientation, $\chi = 29^\text{o}$, i.e. the MRSSP is close to the $(\bar{2}11)$ plane sheared in the antitwinning direction. In simple shear, the Peierls stress would be significantly larger than when the MRSSP is the $(\bar{1}01)$ plane because of the twinning/antitwinning asymmetry. However, traction along orientation $B$ also involves non-glide stresses, with a ratio $NGS/RSS = 0.23$, which reduces the Peierls stress and counterbalances the effect of the twinning/antwinning asymmetry. As a result and as summarized in Table \ref{table}, Peierls stresses with orientation $B$ are close or slightly smaller than in simple shear.

\subsection{Kinked dislocation quantum Peierls stress}

\begin{figure}
\begin{center}
\includegraphics[width=14cm]{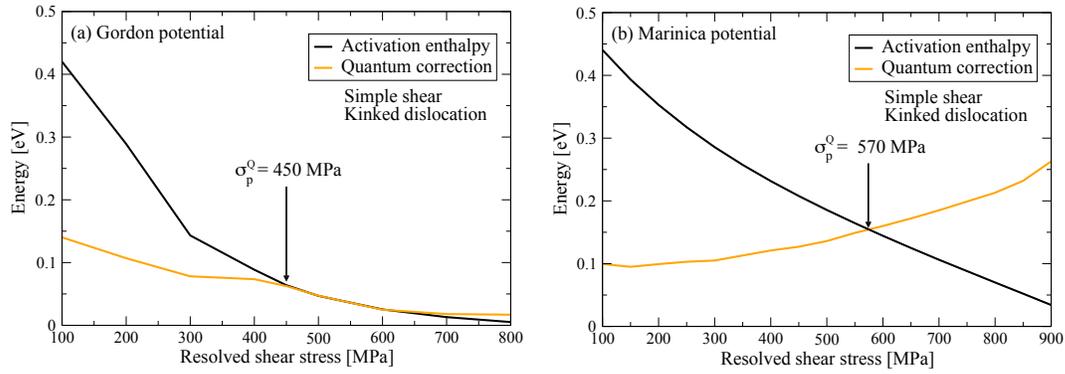}
\caption{\label{fig:3DPeierls}Activation enthalpy and quantum correction of a kinked dislocation loaded in simple shear for (a) Gordon and (b) Marinica potentials. Diagonalizations were performed in cylinders of radius $R = 20 \AA$.}
\end{center}
\end{figure}

\begin{figure}
\begin{center}
\includegraphics[width=14cm]{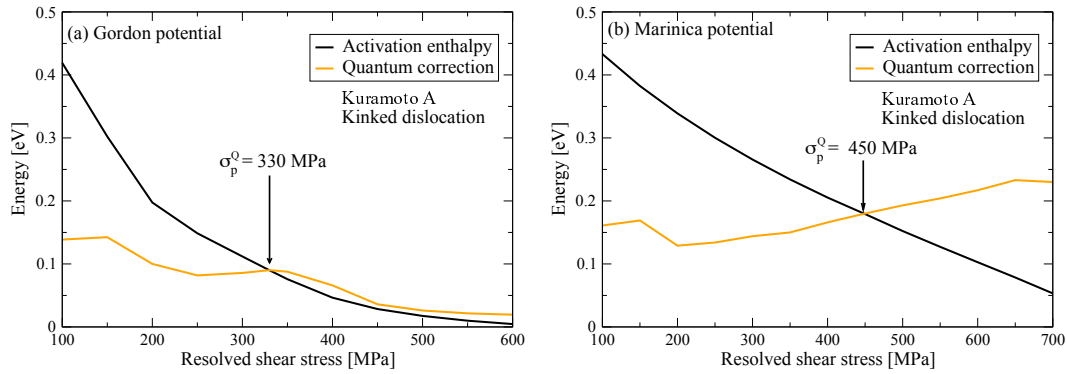}
\caption{\label{fig:3DKuramoto}Activation enthalpy and quantum correction of a kinked dislocation loaded along orientation $A$ for (a) Gordon and (b) Marinica potentials.}
\end{center}
\end{figure}

Calculations on kinked dislocations are shown in Fig. \ref{fig:3DPeierls} in simple shear and in Fig. \ref{fig:3DKuramoto} for orientation $A$ (orientation $B$ is similar to the other curves and is not shown here). In all cases, the quantum correction is large, on the order of 0.1 eV, a significant fraction of the kink-pair activation energy, $\sim$ 0.45 eV.

Fig. \ref{fig:3DPeierls}(a) with Gordon potential in simple shear is similar to that reported in Ref. \cite{proville-natmat2012} with the same quantum Peierls stress, 450 MPa. The quantum Peierls stress in simple shear with Marinica potential is slightly higher, 570 MPa. As discussed in Section \ref{sec:mode}, the quantum correction unexpectedly increases with applied stress with Marinica potential, while it decreases gradually with Gordon potential. The curves obtained along orientation $A$ exhibit the same trends, with quantum Peierls stresses significantly smaller than in simple shear, 330 MPa and 450 MPa for Gordon and Marinica potentials respectively. For orientation $B$ (see Table \ref{table}), the Peierls stress is close to that obtained in simple shear, because of the balancing effects of shearing the crystal in the antitwinning direction and including non-glide stresses, as for the straight dislocation.

\section{Conclusions}

\begin{table}
\begin{center}
\caption{\label{table}Comparison between classical and quantum Peierls stresses obtained with Gordon and Marinica potentials in simple shear and orientations $A$ and $B$ of Kuramoto \etal \cite{kuramoto-pm1979}. Experimental values are from the latter reference.}
\begin{tabular}{lclclclclclc}
\hline
Gordon&$\sigma_P^C$&$\sigma_P^Q(straight)$&$\sigma_P^Q(kinked)$& Exp. \\
\hline
Simple shear &  912 & 330 & 450 & \\
Orientation $A$ &  650  & 200 & 330 & 375 \\
Orientation $B$ &  1000  & 290 & 500 & 450 \\
\hline
Marinica&$\sigma_P^C$ & $\sigma_P^Q(straight)$ & $\sigma_P^Q(kinked)$ & Exp. \\
\hline
Simple shear &  1000 & 410 & 570 & \\
Orientation $A$ &  900  & 330 & 450 & 375 \\
Orientation $B$ &  980  & 360 & 500 & 450 \\
\hline
\end{tabular}
\end{center}
\end{table}

\subsection{Limitations of EAM potentials}

Comparing the predictions of the two EAM potentials, we see that while the kink-pair formation enthalpy behaves  similarly with both potentials, the quantum correction is very different: with Gordon potential, the correction is mostly due to low-frequency modes and decreases with applied stress while with Marinica potential, the correction has a large component arising from high-energy modes and increases with applied stress.

It remains that with both potentials, the correction represents a significant fraction of the kink-pair formation energy, leading in both cases to quantum Peierls stresses significantly smaller than their classical counterparts. This work thus confirms the large role played by quantum effects on the glide of dislocations at low temperatures.

To be more quantitative, ab initio calculations would be helpful and can technically be done, at least for a straight dislocation since Fig. \ref{fig:cylinder2D} shows that the cylinder radius can be decreased down to 15 $\AA$, which contains only 148 atoms. The calculations remain computationally expensive because the Hessian matrix in ab initio has to be constructed by finite differences but the calculations could confirm two important points:  the magnitude of the quantum correction compared to the Peierls barrier and the dependence of the quantum correction on mode frequency to identify more precisely its origin. 

\subsection{Quantum Peierls stresses of straight and kinked dislocations}

The calculations summarized in Tab. \ref{table} show that for both EAM potentials the quantum Peierls stress is smaller for a straight than for a kinked dislocation. Quantum effects are thus stronger in the former case, which could be expected since the full dislocation is brought up the Peierls barrier, implying a larger variation of the normal modes of the crystal than in a simulation cell where only the kink regions are not at the bottom of a Peierls valley. 

The straight dislocation quantum Peierls stress may be very low, down to 200 MPa for Gordon potential with orientation $A$, which is very small compared to the experimental value with this orientation, i.e. 375 MPa. Calculations for a straight dislocation thus probably overestimate the quantum effect but are nevertheless useful because they provide a lower limit for the quantum Peierls stress.

And this raises an important question: what is the critical state at the quantum Peierls stress? In the calculations performed here, we employ a strong approximation by assuming that the atomic configuration of the activated state is that predicted by classical NEB calculations. 
This is probably a good approximation when the barrier is large, but near the quantum Peierls stress, the barrier is low and it would seem physical to expect that at the instability the dislocation becomes straight in the activated state, as near the classical Peierls stress. 
It would be very interesting to include the zero-point energy directly in the NEB calculations, or in other words, to consider the quantum free energy of the system rather than just the potential energy. This was recently done in the classical regime to study the high-temperature athermal limit \cite{gilbert-prl2013}. An extension to the quantum regime \cite{schenter-jcp1994} would be necessary to investigate the low-temperature limit.

 \subsection{Non-glide stresses}

This work is the first study where the effect of non-glide stresses is investigated on the full kink-pair formation enthalpy and not just on the Peierls stress. 
The result is that, when comparing Figs. \ref{fig:2DPeierls}, \ref{fig:2DKuramoto}, \ref{fig:3DPeierls} and \ref{fig:3DKuramoto}, non-glide stresses do not alter the shape of the enthalpy nor of the quantum correction, but only rescale the stresses, leading to a systematic decrease of the Peierls stress in traction compared to simple shear.

Non-glide stresses have a strong effect when comparing the Peierls stresses in simple shear and along orientation $A$, both in the quantum and classical calculations. The effect is more pronounced with Gordon potential, where the decrease is on the order of 30 $\%$ while it is closer to 20 $\%$ with Marinica potential. With orientation $B$, there is a balance between shearing along a $(\bar{2}11)$ plane in antitwinning direction and adding non-glide stresses, such that the quantum Peierls stresses in this orientation are almost the same or slightly below those in simple shear.

In all cases, including non-glide stresses brings the quantum Peierls stresses even closer to the experimental values, in particular with Gordon potential where, starting from the reference value of 912 MPa obtained in simple shear without quantum correction, the quantum Peierls stresses are 300 and 500 MPa for orientations $A$ and $B$, compared to 375 and 450 MPa experimentally. Simple shear is thus the most straightforward and most commonly used loading condition but it strongly overestimates the critical resolved shear stress compared to experimental conditions. Including non-glide stresses is thus essential to compare simulations with experiments.

\section*{Acknowledgments}
The authors wish to thank P. Gumbsch for suggesting to include non-glide stresses in the quantum calculations.

\section*{References}
\bibliographystyle{unsrt}
%\bibliography{PEIERLS}

\end{document}